\documentclass[twocolumn,prl,showpacs,floatfix]{revtex4}

\usepackage{graphicx}
\usepackage{dcolumn}
\usepackage{bm}
\usepackage{amsmath}

\begin{document}

\title{A fully quantal molecular description for the spectra of bosons and fermions 
in the lowest Landau level} 

\author{Constantine Yannouleas and Uzi Landman}

\affiliation{School of Physics, Georgia Institute of Technology,
             Atlanta, Georgia 30332-0430}

\date{19 February 2010; An extended version was published as Phys. Rev. A {\bf 81}, 023609 (2010); see arXiv:1001.1090}

\begin{abstract}
Through the introduction of a class of appropriate translationally invariant
trial wave functions, we show that the strong correlations in the lowest Landau 
level (LLL) reflect in finite systems the emergence of intrinsic point-group 
symmetries associated with rotations and vibrations of molecules formed through 
particle localization. This quantal molecular description is universal, being 
valid for both bosons and fermions, for both the yrast and excited states of the
LLL spectra, and for both low and high angular momenta. "Quantum-fluid" physical
pictures associated with Jastrow-type trial functions are shown to be reducible
to the molecular description introduced in this paper.
\end{abstract}


\maketitle

{\it Motivation.\/} $-$ Following the discovery \cite{tsui82} of the fractional 
quantum Hall effect (FQHE) in two-dimensional (2D) semiconductor 
heterostructures under high magnetic fields ($B$), the description of 
strongly correlated electrons in the lowest Landau level (LLL) developed 
into a major branch of theoretical condensed matter physics \cite{laug8399,
trug85,moor91,ston92,oakn95,jainbook,yann02,yann07,jeon0407}.

Most recently, the burgeoning fields of semiconductor \cite{yann07} 
quantum dots and rapidly rotating trapped ultracold neutral gases 
\cite{mott99,gunn00,pape01,ueda01,popp04,barb06,baks07,vief08,coop08,geme08,
muell08,gunn08}
have generated significant interest pertaining strongly correlated states in 
the lowest Landau level.
Furthermore, it is anticipated that small (and/or mesoscopic) assemblies of 
ultracold bosonic atoms will become technically available in the near future 
\cite{geme08,muell08} and that they will provide an excellent vehicle 
\cite{popp04,barb06,coop08,geme08,muell08,gunn08} 
for experimentally reaching exotic phases and for testing the rich variety of 
proposed LLL trial wave functions, including the Jastrow-Laughlin (JL) 
\cite{laug8399}, composite-fermion (CF) \cite{jainbook}, Moore-Read \cite{moor91}, 
and rotating Wigner molecule (REM) \cite{yann02,yann07} trial functions. 

A universal physical description of the full LLL spectra (including both yrast 
\cite{note3} and all excited states), however, is still missing. To remedy this,
a unified theory for the LLL spectra of a small number of particles valid for 
both statistics (i.e., for both bosons and fermions) is introduced in this 
paper. The LLL spectra are shown to be 
associated with {\it fully quantal\/} \cite{note5} and strongly 
correlated ro-vibrational molecular (RVM) states, i.e., with (analytic) trial 
functions describing vibrational excitations relative to the set of special
yrast states that exhibit enhanced stability and magic angular momenta, and are
referred to as cusp states. The cusp states, important as they are, represent 
only a small fraction of the LLL spectrum. The molecular trial functions 
associated with them are purely rotational (i.e., vibrationless) 
and were introduced for the case of electrons in Ref.\ \cite{yann02} under the 
name rotating electron molecules (REMs). The corresponding {\it analytic\/} 
bosonic trial functions for cusp states [called rotating boson molecules (RBMs)]
are introduced in this paper; see Eq.\ (\ref{rbm1}).

It is remarkable that the numerical results of the present theory were found
in all tested cases to agree within machine precision with exact-diagonalization 
(EXD) results, including energies, wave functions, and overlaps. This numerical
behavior points toward a deeper mathematical finding, i.e., that the RVM trial
functions for both statistics provide a {\it correlated\/} basis (see below) 
that spans the translationally invariant (TI) subspace \cite{trug85} of the LLL 
spectrum. An uncorrelated basis, without physical meaning, built out of 
products of elementary symmetric polynomials is also known to span the (bosonic) 
TI subspace \cite{note4}. We are, however, unaware of any other correlated 
functions which span this subspace. Indeed, although
the Jastrow-Laughlin function is translationally invariant, its quasi-hole
and quasi-electron excitations are not \cite{trug85}. Similarly, the compact
composite-fermion trial functions are translationally invariant \cite{vief08}, 
but the CF excitations which are needed to complete the CF basis are 
not \cite{jeon0407}. The shortcoming of the above well known correlated LLL 
theories to satisfy fundamental symmetries of 
the many-body Hamiltonian represents an unsatisfactory state of affairs, and the
present paper provides a remedy to this effect. At the same time a different 
physical picture is being established: that is, the RVM functions and 
corresponding molecular point-group symmetries are superior in expressing and 
interpreting the emergent many-body correlations of the LLL states.

{\it Theory.\/} $-$ The RVM functions have the general form
(within a normalization constant):
\begin{equation}
\Phi^{\text{RXM}}_{\cal L}(n_1,n_2) Q_\lambda^m |0>, 
\label{mol_trial_wf}
\end{equation}
where $(n_1,n_2)$ indicates the molecular configuration (here we consider 
two concentric rings) of point-like particles with $n_1$ ($n_2$) 
particles in the first (second) ring. The particles on each ring form 
regular polygons. The index RXM stands 
for either REM, i.e., a rotating electron molecule, or RBM, i.e., a rotating 
boson molecule. $\Phi^{\text{RXM}}_{\cal L}(n_1,n_2)$ 
alone describes pure molecular rotations associated with magic angular momenta 
${\cal L}={\cal L}_0+n_1k_1 + n_2k_2$, with $k_1$, $k_2$ being nonnegative 
integers; ${\cal L}_0=N(N-1)/2$ for electrons and ${\cal L}_0=0$ for bosons. 
The product in Eq.\ (\ref{mol_trial_wf}) combines rotations 
with vibrational excitations, the latter being 
denoted by $Q_\lambda^m$, with $\lambda$ being an angular momentum; the 
superscript denotes raising to a power $m$. Both $\Phi^{\text{RXM}}_{\cal L}$ 
and $Q_\lambda^m$ are homogeneous polynomials of the complex particle coordinates 
$z_1,z_2,\ldots,z_N$, of order ${\cal L}$ and $\lambda m$, respectively.
The total angular momentum $L={\cal L}+\lambda m$. $Q_\lambda^m$ is always 
symmetric in these variables; $\Phi^{\text{RXM}}_{\cal L}$ is antisymmetric 
(symmetric) for fermions (bosons). $|0> = \prod_{i=1}^N \exp[-z_iz_i^*/2]$; this
product of Gaussians will be omitted henceforth.

The analytic expressions for the $\Phi^{\text{REM}}_{\cal L}$ (for fully 
polarized electrons) were derived in Ref.\ \cite{yann02} employing a 
two-step method: (i) First a single Slater {\it determinant\/} [that breaks the 
rotational (circular) symmetry] was constructed using displaced Gaussians as 
electronic orbitals, i.e.,  
\begin{eqnarray}
u(z,Z_j) = \frac{1}{\sqrt{\pi}} \exp[-|z-Z_j|^2/2] \exp[-i (xY_j+yX_j)].
\label{gaus}
\end{eqnarray}
The phase factor is due to the gauge invariance. $z \equiv x-i y$,
and all lengths are in dimensionless units of ${l_B}\sqrt{2}$, with 
the magnetic length $l_B=\sqrt{\hbar /(m_e \omega_c)}$;  
$\omega_c = eB/(m_ec)$ is the cyclotron frequency. The centers 
$Z_j \equiv X_j+i Y_j$, $j=1,2,\ldots,N$ of the Gaussians are the vertices of 
the regular polygons in the $(n_1,n_2)$ geometric arrangement. (ii) A subsequent
step of symmetry restoration was performed using the projection operator
${\cal P}({\cal L})=\frac{1}{2\pi} \int_0^{2 \pi} d\gamma 
e^{i\gamma(\hat{L}-{\cal L})}$, where $\hat{L}=\sum_{i=1}^N \hat{l}_i$ is the 
total angular momentum operator; this yielded trial wave functions with good 
total angular momenta ${\cal L}$ \cite{yann02,yann07}.

Analytic expressions for the $\Phi^{\text{RBM}}_{\cal L}$ (for spinless bosons)
can also be derived using the two-step method. Naturally, in the first step
one constructs a {\it permanent} out of the orbitals of Eq.\ (\ref{gaus}); one 
also uses the equivalence $\omega_c \rightarrow 2\Omega$ between the cyclotron 
frequency $\omega_c$ (electrons) and the rotational frequency $\Omega$ (bosons) 
\cite{yann07}. 
Here we present as an illustrative example the simpler case of $N=3$ (and $N=4$ 
in the Appendix) bosons having a $(0,N)$ one-ring 
molecular configuration. One has (within a normalization constant)  
\begin{equation}
\Phi^{\text{RBM}}_{\cal L}(0,3) = 
\sum_{0 \leq l_1 \leq l_2 \leq l_3}^{l_1+l_2+l_3={\cal L}}
C(l_1,l_2,l_3) \;{\text{Perm}} [z_1^{l_1}, z_2^{l_2}, z_3^{l_3}],
\label{rbm1}
\end{equation}
where the symbol "Perm" denotes a permanent with elements 
$z_i^{l_j}$, $i,j=1,2,3$; only the diagonal elements are shown in Eq.\ 
(\ref{rbm1}). The coefficients were found to be:
\begin{eqnarray}
C(l_1,l_2,l_3) &=& \left(\prod_{i=1}^3 l_i! \right)^{-1} 
\left(\prod_{k=1}^M p_k! \right)^{-1}
\nonumber \\
&\times& \left( \sum_{1 \leq i < j \leq 3} 
\cos \left[\frac{2\pi(l_i-l_j)}{3} \right] \right),
\label{rbm2}
\end{eqnarray} 
where $1 \leq M \leq 3$ denotes the number of different
indices in the triad $(l_1,l_2,l_3)$ and the $p_k$'s are the 
multiplicities of each one of the different indices. For example, 
for (1,1,4), one has $M=2$ and $p_1=2$, $p_2=1$. 

The $\Phi^{\text{REM}}_{\cal L}$ expressions for electrons in a $(0,N)$ 
or a $(1,N-1)$ configuration are given by Eqs.\ (6.2) and (6.4) of Ref.\ 
\cite{yann07}, respectively. For electrons (1) $M=N$ in all instances and (2) a 
{\it product of sine\/} terms replaces the {\it sum of cosine\/} terms appearing 
in Eq.\ (\ref{rbm2}). 

We note that $\Phi^{\text{RXM}}_{\cal L}(n_1,n_2) =0$ for both bosons and 
electrons when ${\cal L} \neq {\cal L}_0 + n_1 k_1 + n_2 k_2$. This selection 
rule follows directly from the point group symmetries of the $(n_1,n_2)$ 
molecular configurations.

The vibrational excitations $Q_\lambda$ are given by the same expression for
both bosons and electrons, namely, by the symmetric polynomials:
\begin{equation}
Q_\lambda = \sum_{i=1}^N (z_i-z_c)^\lambda,
\label{ql} 
\end{equation}
where $z_c=(1/N)\sum_{i=1}^N z_i$ is the coordinate of the center of mass and 
$\lambda>1$ is a prime number. Vibrational excitations of a similar form, i.e., 
$\tilde{Q}_\lambda=\sum_{i=1}^N z_i^\lambda$ (and certain other variations), 
have been used earlier to approximate {\it part\/} of the LLL spectra. Such 
earlier endeavors provided valuable insights, but overall they remained 
inconclusive; e.g., for electrons in the neighborhood of the maximum density 
droplet [with fractional filling $\nu=1$ ($\nu=N(N-1)/2{\cal L}$)], 
see Refs.\ \cite{ston92} and \cite{oakn95}, and 
for bosons in the range $0 \leq L \leq N$, see Refs.\ 
\cite{mott99,ueda01,pape01}. 

The advantage of $Q_\lambda$ is that it is translationally invariant 
(TI) \cite{trug85,pape01}, a property shared with both  
$\Phi^{\text{RBM}}_{\cal L}$ and $\Phi^{\text{REM}}_{\cal L}$.
In the following, we will discuss illustrative cases, which will show that
the RVM functions of Eq.\ (\ref{mol_trial_wf}) provide a correlated
basis (RVM basis) that spans the TI subspace \cite{trug85,pape01,vief08} of 
{\it nonspurious\/} states in the LLL spectra. The dimension 
$D^{\text{TI}}(L)$ of the RVM-diagonalization space (using the RVM basis) is 
much smaller than the dimension $D^{\text{EXD}}(L)$ of the 
exact-diagonalization (EXD) \cite{yann07} space spanned by uncorrelated 
determinants ${\text{Det}} [z_1^{l_1}, \ldots, z_N^{l_N}]$ or permanents 
${\text{Perm}} [z_1^{l_1}, \ldots, z_N^{l_N}]$ formed with Darwin-Fock
orbitals. The remaining $D^{\text{EXD}}(L) - D^{\text{TI}}(L)$ states are 
{\it spurious\/} center-of-mass excitations (generated by applying 
$\tilde{Q}_1^m$) whose energies coincide with those 
appearing at all the other smaller angular momenta \cite{trug85}. Thus 
$D^{\text{TI}}(L)=D^{\text{EXD}}(L)- D^{\text{EXD}}(L-1)$; see TABLE 
\ref{ene_bos_np3}. 
 
\begin{table*}[t] 
\caption{\label{ene_bos_np3}%
LLL spectra of three spinless bosons interacting via a repulsive contact 
interaction $g \delta(z_i-z_j)$. 2nd column: Dimensions of the EXD and 
the nonspurious TI (in parenthesis) spaces (the EXD space is spanned by 
uncorrelated permanents of Darwin-Fock orbitals). 
4th to 6th columns: Matrix elements [in units of $g/(\pi \Lambda^2)$, 
$\Lambda=\sqrt{\hbar/(m\Omega)}$] of the contact interaction between the 
correlated RVM states $\{k,m \}$ [see Eq.\ (\ref{wfbos3})]. The total angular 
momentum $L=3k+2m$. Last three columns: Energy eigenvalues from the RVM 
diagonalization of the associated matrix of dimension $D^{\text{TI}}(L)$. There 
is no nonspurious state with $L=1$. The full EXD spectrum at a given $L$ is 
constructed by including, in addition to the listed TI eigenvalues 
[$D^{\text{TI}}(L)$ in number], all the energies associated with angular 
momenta smaller than $L$. An integer in square brackets indicates the energy 
ordering in the full EXD spectrum (including both spurious and TI states).
Seven decimal digits are displayed, but the energy eigenvalues from the RVM 
diagonalization agree with the corresponding EXD-TI ones within machine 
precision.}
\begin{ruledtabular}
\begin{tabular}{rllllllll}
$L$ & $D^{\text{EXD}}(D^{\text{TI}})$ & $\{k,m\}$ & 
\multicolumn{3}{l}{Matrix elements} & 
\multicolumn{3}{l}{Energy eigenvalues (RVM diag. or EXD-TI)} \\ \hline
0 & 1(1)  & \{0,0\} & 1.5000000  & 
  ~~~~      &  ~~~~ & 1.5000000[1] & ~~~~ & ~~~~\\
2 & 2(1) & \{0,1\} & 0.7500000  &
  ~~~~      & ~~~~ & 0.7500000[1] & ~~~~ & ~~~~\\
3 & 3(1) & \{1,0\} & 0.3750000  &
  ~~~~      & ~~~~ & 0.3750000[1] & ~~~~ & ~~~~\\
4 & 4(1) & \{0,2\} & 0.5625000  &
  ~~~~      & ~~~~ & 0.5625000[2] & ~~~~ & ~~~~\\
5 & 5(1) & \{1,1\} & 0.4687500  & 
  ~~~~      & ~~~~ & 0.4687500[2] & ~~~~ & ~~~~\\
6 & 7(2) & \{2,0\} & 0.0468750  & 
 0.1482318  &  ~~~~ &  ~~~~       & ~~~~ & ~~~~\\
~~~   & ~ & \{0,3\} & 0.1482318  & 0.4687500   & ~~~~ & 0.0000000[1] & 
0.5156250[4] & ~~~~\\
7 & 8(1) & \{1,2\} & 0.4921875  & 
  ~~~~      & ~~~~ & 0.4921875[4] & ~~~~ & ~~~~\\
8 & 10(2) & \{2,1\} & 0.0937500  & 
0.1960922   &~~~~ &  ~~~~         & ~~~~ & ~~~~\\
~~~   & ~ & \{0,4\} & 0.1960922  & 0.4101562  & ~~~~ & 0.0000000 & 
0.5039062[6] & ~~~~\\
12 & 19(3) & \{4,0\} & 7.3242187$\times 10^{-4}$ & 1.0863572$\times 10^{-2}$ & 
1.5742811$\times 10^{-2}$ & ~~~~~~  & ~~~~~~  & ~~~~~ \\
~~~ & ~~~ & \{2,3\} & 1.0863572$\times 10^{-2}$ & 0.1611328 & 
0.2335036 & ~~~~~~ & ~~~~~  & ~~~~~ \\
~~~ & ~~~ & \{0,6\} & 1.5742811$\times 10^{-2}$ & 0.2335036 & 
0.3383789 & 0.0000000  & 0.0000000  & 0.5002441[13] \\
\end{tabular}
\end{ruledtabular}
\end{table*}

{\it Three spinless bosons\/} $-$ Only the $(0,3)$ molecular 
configuration and the dipolar $\lambda=2$ vibrations are at play
(as checked numerically), i.e.,
the full TI spectra at any $L$ are spanned by the wave functions 
\begin{equation}
\Phi^{\text{RBM}}_{3k}(0,3) Q_2^m \Rightarrow \{k,m\},
\label{wfbos3}
\end{equation} 
with $k,m=0,1,2,\ldots$, and $L=3k+2m$; these states 
are always orthogonal. This represents a remarkable analogy with the case of 
$N=3$ electrons (see below). 

TABLE \ref{ene_bos_np3} provides the systematics of the molecular
description for the beginning ($0 \leq L \leq 12$) of the LLL spectrum. 
There are several cases when the TI
subspace has dimension one and the exact solution coincides with a single
$\{k,m\}$ state. For $L=0$ the exact solution coincides with 
$\Phi^{\text{RBM}}_0=1$ ($Q_\lambda^0=1$); 
this is the only case when an LLL state has a Gross-Pitaevskii form, i.e., it is
a single permanent [see $|0\rangle$ in Eq.\ (\ref{mol_trial_wf})].
For $L=2$, we found $\Phi^{\text{exact}}_{[1]} \propto Q_2$ (for the index 
$[i]$, see caption of TABLE \ref{ene_bos_np3}), and since [see Eq.\ (\ref{ql})]
$Q_2 \propto(z_1-z_c)(z_2-z_c)+(z_1-z_c)(z_3-z_c)+(z_2-z_c)(z_3-z_c)$, this 
result agrees with the findings of Refs.\ \cite{smit00,pape01} concerning 
ground states of bosons in the range $0 \leq L \leq N$. For $L=3$, one finds
$\Phi^{\text{exact}}_{[1]} \propto \Phi^{\text{RBM}}_3$. Since 
$\Phi^{\text{RBM}}_3 \propto(z_1-z_c)(z_2-z_c)(z_3-z_c)$ [see Eq. (\ref{rbm1})], 
this result agrees again with the findings of Refs.\ \cite{smit00,pape01}. 
For $L=5$, the single nonspurious state is an excited one, 
$\Phi^{\text{exact}}_{[2]} \propto \Phi^{\text{RBM}}_3 Q_2$.
For $L=6$, the ground-state is found to be $\Phi^{\text{exact}}_{[1]} \propto
-160\Phi^{\text{RBM}}_6/9  + Q_2^3/4=(z_1-z_2)^2(z_1-z_3)^2(z_2-z_3)^2$, i.e., 
the bosonic Laughlin function for $\nu=1/2$ is equivalent to an RBM state that 
incorporates vibrational correlations. For $L \geq N(N-1)$ (i.e.,
$\nu \leq 1/2$), the EXD yrast 
energies equal zero, and with increasing $L$ the degeneracy of the zero-energy 
states for a given $L$ increases. It is important that this nontrivial behavior 
is reproduced faithfully by the present method (see TABLE \ref{ene_bos_np3}).

{\it Three electrons\/} $-$ Although unrecognized, the solution of the
problem of three spin-polarized electrons in the LLL using molecular trial 
functions has been presented in Ref.\ \cite{laug83.2}. 
Indeed, the wave functions in Jacobi coordinates in Eq. (18) of Ref.\ 
\cite{laug83.2} are \cite{yann02} precisely of the form 
$\Phi^{\text{REM}}_{3k} Q_2^m$, as can be checked after transforming back to 
cartesian coordinates. It is noteworthy that Laughlin did not present 
molecular trial functions for electrons with $N >3$, or for bosons for any $N$. 
This is done in the present paper. 

\begin{table}[b] 
\caption{\label{exp_coeff}%
$N=4$ LLL electrons with $L=18$: Expansion coefficients in the RVM basis 
(labelled by the $|i>$'s) for the three lowest-in-energy EXD-TI states (labelled
[1], [2], [4]; see the Appendix). The 4th column gives the RVM expansion 
coefficients of the corresponding Jastrow-Laughlin expression. 
}
\begin{ruledtabular}
\begin{tabular}{ccccc}
RVM & EXD-TI [1] & EXD-TI [2] & EXD-TI [4] & JL \\
\hline
$|1>$ & \underline{0.9294} & -0.3430 & 0.0903 & 0.8403 \\ 
$|2>$ & -0.1188 & -0.0693 & \underline{0.8930} & -0.1086 \\ 
$|3>$ & 0.0067  &  0.0382 & -0.2596 & 0.0076 \\ 
$|4>$ & 0.0137  &  0.0191 & -0.0968 & 0.0395 \\ 
$|5>$ & 0.2540  &  \underline{0.8486} & 0.1519 & 0.4029  \\ 
$|6>$ & 0.0211  &  0.0283 & 0.3097 & 0.0616 \\ 
$|7>$ & -0.2387 & -0.3935 & 0.0877 & -0.3380 \\ 
\end{tabular}
\end{ruledtabular}
\end{table}

{\it Four electrons\/} $-$ For $N=4$ spin-polarized electrons, one needs to
consider two distinct molecular configurations, i.e., $(0,4)$ and $(1,3)$.
Vibrations with $\lambda \geq 2$ must also be considered. In this case the 
RVM states are not always orthogonal, and the Gram-Schmidt orthogonalization is 
implemented. 

\begin{figure}[t]
\centering\includegraphics[width=8.4cm]{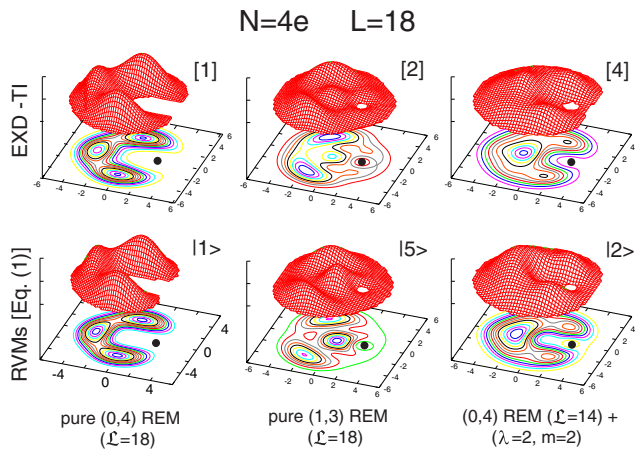}
\caption{CPDs for $N=4$ LLL electrons with $L=18$ ($\nu=1/3$). 
Top row: The three lowest-in-energy EXD-TI states (see the 
Appendix). Bottom row: The RVM trial functions associated with 
the largest expansion coefficients (underlined, see TABLE \ref{exp_coeff}) of 
these three EXD-TI states in the correlated RVM basis. See the text for 
details. The solid dot denotes the fixed point ${\bf r}_0$. Distances in nm. 
}
\label{cpdsexdrvm}
\end{figure}

Of particular interest is the $L=18$ case ($\nu=1/3$) which is considered 
\cite{laug8399} as the  prototype of quantum-liquid states. However, in this case 
we found (see the Appendix) that the exact 
TI solutions are linear superpositions of the following seven RVM states 
[involving both the (0,4) and (1,3) configurations]:
$|1>=\Phi^{\text{REM}}_{18}(0,4)$, $|2>=\Phi^{\text{REM}}_{14}(0,4) Q_2^2$,
$|3>=\Phi^{\text{REM}}_{10}(0,4) Q_2^4$, $|4>=\Phi^{\text{REM}}_{6}(0,4) Q_2^6$,
$|5>=\Phi^{\text{REM}}_{18}(1,3)$, $|6>=\Phi^{\text{REM}}_{12}(1,3) Q_2^3$, and
$|7>=\Phi^{\text{REM}}_{15}(1,3) Q_3$. The expansion coefficients of the three 
lowest-in-energy EXD-TI states (labelled [1], [2], [4]; see the Appendix) 
in this RVM basis are listed in TABLE \ref{exp_coeff}. One sees that for each 
case, one component (underlined) dominates this expansion; this applies 
for both the yrast state (No. [1]) and the two excitations (Nos. [2] and [4]). 
To further illustrate this, we display in Fig.\ \ref{cpdsexdrvm} the 
conditional probability (pair correlation) distributions (CPDs) 
$P({\bf r}, {\bf r}_0)$ (see Eq. (1.1) in Ref.\ \cite{yann07}) for these three 
EXD-TI states (top row) and for the RVM functions (bottom row) corresponding to
the dominant expansion coefficients. The similarity of the CPDs in each column 
is noticeable and demonstrates that the single RVM functions capture 
the essence of many-body correlations in the EXD-TI states. Full quantitative 
agreement (within machine precision) in total energies can be reached by taking 
into consideration all seven RVM basis states (see the Appendix; also for 
other $L$'s, including $L=30$ with 19 RVM basis states). Naturally, a smaller 
number of RVM states yields intermediate degrees of high-quality quantitative 
agreement.      

The celebrated JL ansatz 
$\prod_{1\leq i<j \leq N} (z_i-z_j)^{2p+1}$ has been given exclusively an
interpretation of a quantum-fluid state \cite{laug8399,jainbook}. However, since 
the RVM functions span the TI subspace, it follows that any TI trial function
(including the JL ansatz above and the compact CF states) can be expanded
in the RVM basis. As an example, we give in TABLE \ref{exp_coeff} (4th column) 
the RVM expansion of the JL state for $N=4e$ and $L=18$. One sees that, 
compared to the EXD yrast state (1st column), the relative weight of the pure 
(0,4) REM (denoted by $|1>$) is reduced, while the weights of higher-in-energy
vibrational excitations are enhanced. In this context, the liquid 
characteristics are due to the stronger weight of molecular vibrations which 
diminish the rigidity of the molecule.   

{\it Conclusions.\/} $-$ 
The many-body Hilbert space corresponding to the translationally 
invariant part of the LLL spectra of small systems (whether fermions or bosons,
and for both low and high angular momenta) is spanned by the 
RVM trial functions introduced in Eq.\ (\ref{mol_trial_wf}). 
The yrast and excited states for both short- and long-range interactions can 
always be expressed as linear superpositions of these RVM functions. Thus the 
nature of strong correlations in the LLL reflects the emergence of 
intrinsic point-group symmetries associated with rotations and vibrations of 
molecules formed through particle localization.
We stress the validity of the molecular theory for {\it low\/} angular momenta, 
where "quantum-liquid" physical pictures \cite{laug8399,jainbook,coop08}
have been thought to apply exclusively. Our analysis suggests that liquid-type 
pictures, associated with translationally invariant trial functions (e.g., 
the JL and compact CF functions), are reducible to a description in terms of an 
excited rotating/vibrating quantal molecule. 

Work supported by the U.S. DOE (FG05-86ER45234).\\
~~~~~~\\
~~~~~~\\

\begin{center}
APPENDIX
\end{center}
~~~~~~\\


\begin{table*}[t] 
\caption{\label{ene_ferm_np4}%
LLL spectra of four spin-polarized electrons interacting via the Coulomb 
repulsion $e^2/(\kappa r_{ij})$ . Second column: Dimensions of the full EXD 
and the nonspurious TI (in parenthesis) spaces (the EXD space is spanned by 
uncorrelated determinants of Darwin-Fock orbitals). Last three columns: Energy 
eigenvalues [in units of $e^2/(\kappa l_B)$] from the diagonalization of the 
Coulomb interaction in the TI subspace spanned by the trial functions 
$\Phi^{\text{REM}}_{6+4k}(0,4) Q_\lambda^m$ and 
$\Phi^{\text{REM}}_{6+3k}(1,3) Q_\lambda^m$ (RVM digonalization). 
Third to sixth columns: the molecular configurations $(n_1,n_2)$ and the 
quantum numbers $k$, $\lambda$ and $m$ are indicated within brackets. 
There is no nonspurious state with $L=7$. The full EXD spectrum at a given $L$ 
is constructed by including, in addition to the listed TI energy 
eigenvalues [$D^{\text{TI}}(L)$ in number], all the energies associated with 
angular momenta smaller than $L$. An integer in square brackets indicates the 
energy ordering in the full EXD spectrum (including both spurious and TI 
states), with [1] denoting an yrast state. Eight decimal digits are displayed, 
but the energy eigenvalues from the RVM diagonalization agree with the 
corresponding EXD-TI ones within machine precision.}
\begin{ruledtabular}
\begin{tabular}{llllll}
$L$ & $D^{\text{EXD}}$($D^{\text{TI}}$)& $[(n_1,n_2)\{ k, \lambda, m \}]$ &
\multicolumn{3}{l}{Energy eigenvalues (RVM diag. or EXD-TI)} \\
\hline
6 & 1(1) & [(0,4)\{0,$\lambda$,0\}] & 
2.22725097[1]  &   ~~~~      & ~~~~~ \\
8 & 2(1) & [(0,4)\{0,2,1\}] & 
2.09240211[1] &   ~~~~      & ~~~~~~ \\
9 & 3(1) & [(1,3)\{1,$\lambda$,0\}] &
 1.93480798[1] &   ~~~~      & ~~~~ \\
10 & 5(2) & [(0,4)\{1,$\lambda$,0\}] [(0,4)\{0,2,2\}] & 
 1.78508849[1]  &  1.97809256[3] & ~~~~ \\
11 & 6(1) & [(1,3)\{1,2,1\}] & 1.86157215[2]  & 
  ~~~~      & ~~~~~ \\
12 & 9(3) &  [(0,4)\{1,2,1\}] [(0,4)\{0,2,3\}] [(1,3)\{2,$\lambda$,0\}] &
 1.68518201[1] & 1.76757420[2]  & 1.88068652[5] \\
13 & 11(2) & [(1,3)\{1,2,2\}] [(0,4)\{1,3,1\}]  & 1.64156849[1]  & 
 1.79962234[5] & ~~~~ \\
14 & 15(4) & [(0,4)\{2,$\lambda$,0\}] [(0,4)\{1,2,2\}] [(0,4)\{0,2,4\}] & 
 1.50065835[1]  & 1.63572496[2]   &  1.72910626[5]  \\
~~~ & ~~~~ & [(1,3)\{2,2,1\}] & 
 1.79894008[8] & ~~~~~   & ~~~~~~ \\
15 & 18(3) & [(1,3)\{3,$\lambda$,0\}] [(1,3)\{2,3,1\}] [(1,3)\{1,3,2\}]  & 
 1.52704695[2] & 1.62342533[3] & 1.74810279[8] \\
18 & 34(7) & [(0,4)\{3,$\lambda$,0\}] [(0,4)\{2,2,2\}] [(0,4)\{1,2,4\}] & 
 1.30572905[1]  &  1.41507954[2]   &  1.43427543[4] \\
~~~~ & ~~~~ & [(0,4)\{0,2,6\}] [(1,3)\{4,$\lambda$,0\}] [(1,3)\{2,2,3\}]  & 
 1.50366728[8]  &   1.56527615[11]   & 1.63564655[15] \\
~~~~ & ~~~~ & [(1,3)\{3,3,1\}]  & 
 1.68994048[20]  &   ~~~~      & ~~~~~ \\ 
\end{tabular}
\end{ruledtabular}
\end{table*}

{\it RBM analytic expression for $N=4$ bosons\/} $-$
Here we present the case of $N=4$ bosons having a $(0,4)$ one-ring molecular 
configuration. One has (within a normalization constant)  
\begin{eqnarray}
\Phi^{\text{RBM}}_{\cal L}(0,4)&=& \nonumber \\ 
&& \hspace{-3.0cm}
\sum_{0 \leq l_1 \leq l_2 \leq l_3 \leq l_4}^{l_1+l_2+l_3+l_4={\cal L}}
C(l_1,l_2,l_3,l_4) \;{\text{Perm}} [z_1^{l_1}, z_2^{l_2}, z_3^{l_3}, z_4^{l_4}],
\label{rbm22}
\end{eqnarray}
where the symbol ``Perm'' denotes a permanent with elements $z_i^{l_j}$, 
$i,j=1,2,3,4$; only the diagonal elements are shown in Eq.\ (\ref{rbm22}). The 
coefficients were found to be:
\begin{eqnarray}
C(l_1,l_2,l_3,l_4) &=& \left(\prod_{i=1}^4 l_i! \right)^{-1} 
\left(\prod_{k=1}^M p_k! \right)^{-1}
\nonumber \\
&& \hspace{-2.0cm} \times \left( 
\sum_{n=1}^{4!} {\cal P}_n \cos\left[ (3l_1+l_2-l_3-3l_4)\frac{\pi}{4} \right]
\right),
\label{rbm33}
\end{eqnarray} 
where ${\cal P}_n$ is an operator that generates the $n$th permutation
of the boson labels (subscripts of $l_i$'s) 1, 2, 3, and 4. The index $M$ (with 
$1 \leq M \leq 4$) denotes the number of different indices in the tetrad 
$(l_1,l_2,l_3,l_4)$ and the $p_k$'s are the multiplicities of each one of the 
different indices. For example, for (2,2,2,5), one has $M=2$ and 
$p_1=3$, $p_2=1$; for (0,0,0,0), one has $M=1$ and $p_1=4$;  
for (1,2,3,9), one has $M=4$ and $p_1=p_2=p_3=p_4=1$.  

Expressions for any number $N$ of bosons and any molecular configuration 
$(n_1,n_2, \ldots ,n_q)$ can also be derived and will be presented in a
future publication \cite{yann09}. 

{\it LLL spectrum of four polarized electrons\/} $-$
For $N=4$ spin-polarized electrons, one needs to consider rovibrational states 
[see Eq.\ (1) in the main text] for two distinct molecular configurations, i.e., 
$\Phi^{\text{REM}}_{6+4k}(0,4) Q_\lambda^m$ and 
$\Phi^{\text{REM}}_{6+3k}(1,3) Q_\lambda^m$. TABLE \ref{ene_ferm_np4} 
summarizes the quantal molecular description in the start of the 
LLL spectrum ($6 \leq L \leq 15$ and $L=18$). 

In several cases the nonspurious (TI) states are given by a single trial 
function as defined in Eq.\ (1) of the main text. Indeed for $L=9$ 
the yrast state is a pure REM state, i.e., $\Phi^{\text{REM}}_{9}(1,3)$. 
For $L=11$ the single nonspurious state is the first excited state in the full 
spectrum,  coinciding with the molecular vibration 
$\Phi^{\text{REM}}_{9}(1,3)Q_2$. 

Of particular interest is the $L=18$ case; it corresponds to the celebrated
$\nu=1/3$ fractional filling, which is considered \cite{laug8399} as the 
prototype of quantum liquid states. However, in this case we found (see TABLE
\ref{ene_ferm_np4}) that the EXD-TI nonspurious solutions are 
linear superpositions of seven molecular states involving dipole $(\lambda=2)$ 
and octupole $(\lambda=3)$ vibrations relative to both the (0,4) and (1,3) 
configurations. Focusing on the yrast state with $L=18$, we found that its 
largest component is the pure $\Phi^{\text{REM}}_{18}(0,4)$ REM state with a 
0.9294 overlap with the EXD solution (see TABLE II in the main text); the 
contributions of the remaining six states are much smaller, 
but they bring the overlap to precisely unity. Unlike the $\nu=1/2$ case of 
bosons, we stress that the fermionic Jastrow-Laughlin functions at all $\nu$'s 
exhibit less-than-unity overlaps \cite{laug8399,jainbook}.

Of great interest also is the $L=30$ ($\nu=1/5$) case, which in the
composite-fermion picture was found to be susceptible to a competition 
\cite{chan06} between crystalline and liquid orders. However, we found that the 
exact nonspurious states for $L=30$ are actually linear superpositions of the 
following 19 $[=D^{\text{TI}}(L=30)]$ RVM functions: 
$\Phi^{\text{REM}}_{6+4k}(0,4)Q_2^{12-2k}$, with $k=0,1,2,3,4,5,6$; 
$\Phi^{\text{REM}}_{6+3k}(1,3)Q_2^{12-3k/2}$, with $k=2,4,6$;
$\Phi^{\text{REM}}_{6+4k}(0,4)Q_3^{8-4k/3}$, with $k=0,3$; and
$\Phi^{\text{REM}}_{6+3k}(1,3)Q_3^{8-k}$, with $k=2,3,4,5,6,7,8$.
Diagonalization of the Coulomb interaction in the above TI subspace yielded an 
energy 0.25084902 $e^2/(\kappa l_B)$ per electron for the yrast state; this 
value agrees again, within machine precision, with the EXD result. 
The most sophisticated variants of the composite-fermion theory [including 
composite-fermion diagonalization (CFD), composite-fermion crystal (CFC), and 
mixed liquid-CFC states \cite{jainbook,chan06,jeon0407}] fall short in this 
respect. Indeed the following higher energies were 
found \cite{chan06,note23}: 0.250863(6) (CFD), 0.25094(4) (mixed), 0.25101(4) (CFC).
The CFD basis is not translationally invariant \cite{jeon0407}. 
Consequently, to achieve machine-precision accuracy, the CFD will have to be performed
in the larger space of dimension $D^{\text{EXD}}(L=30)=169$.


\begin{thebibliography}{999}
\bibitem{note3}
An yrast state has the lowest enegy at a given $L$.
\bibitem{note4}
Ref.\ \cite{trug85} multiplied the elementary symmetric polynomials by a 
Pauli-exclusion single-determinant factor to generate a basis spanning the 
fermionic TI subspace, but which has no physical picture associated with it.
\bibitem{note23}
Numbers shown in parentheses are statistical uncertainties arising from Monte Carlo 
sampling.
\end{thebibliography}
\end{document}